# Tailoring spin reorientation and magnetic interaction for room-temperature spintronics in Tb-Doped SmFeO₃ single crystal

Mingzhu Xue,[1, 2, a)] Xin Li,[3, a)] Shilei Ding,[4] Qixin Li,[5] Wenhao Di,[5] Anhua Wu,[5] Bin He,[6] Shishen Yan,[6] Wenyun Yang,[7,b)] Jinbo Yang[7]

[1]School of Physics and Astronomy, Beijing Normal University, Beijing 100875, China
[2]Key Laboratory of Multiscale Spin Physics (Ministry of Education), Beijing Normal University, Beijing 100875, China
[3]Department of Physics and Astronomy, University of Nebraska-Lincoln, Lincoln, NE 68588, USA
[4]Department of Materials, ETH Zürich, Zürich 8093, Switzerland
[5]State Key Laboratory of Functional Crystals and Devices, Shanghai Institute of Ceramics, Chinese Academy of Sciences, Shanghai 201899, China
[6]Spintronics Institute, School of Physics and Technology, University of Jinan, Jinan 250022, China
[7]State Key Laboratory for Mesoscopic Physics, School of Physics, Peking University, Beijing 100871, China

a) These authors contributed equally to this work.
b) Authors to whom correspondence should be addressed: yangwenyun@pku.edu.cn

**ABSTRACT**
Selective doping with different $R$-site ions in rare-earth perovskite $R$FeO₃ compounds offers an effective way to achieve atomic-scale tuning of the complex exchange interactions. In this study, the spin reorientation temperature of Tb-doped SmFeO₃ (Sm$_{0.7}$Tb$_{0.3}$FeO₃) single crystal is lowered to approximately 350 K, making it more suitable for room-temperature applications. Notably, the magnetic compensation point is absent at low temperatures, and both $R^{3+}$ and Fe$^{3+}$ ion moments can be fully saturated under high magnetic fields, suggesting that Tb$^{3+}$ doping drives the $R^{3+}$ and Fe$^{3+}$ sublattices toward ferromagnetic coupling. Moreover, the hysteresis loop along the $a$-axis transitions from a double triangle shape below the spin reorientation temperature to a rectangular shape above the spin reorientation temperature, and the nucleation field exhibits a strong dependence on both the measurement temperature and the maximum applied magnetic field. Above results can be explained by a modified two-domain model with mean field correction. The results of magnetic domain measurements indicate that the emergence of the double-triangular hysteresis loop is jointly determined by domain wall motion and the nonlinear response of the parasitic magnetic moment along the $a$-axis. These findings provide valuable insights and new materials for advancing the use of $R$FeO₃ compounds in spintronics.

Rare-earth orthorhombic perovskites $R$FeO₃ have emerged as a promising class of materials in spintronics due to their intriguing magnetic and magneto-optical properties.[1,2] As canted antiferromagnetic materials, $R$FeO₃ contains two magnetic sublattices, Fe$^{3+}$ and $R^{3+}$, and three types of magnetic coupling (Fe$^{3+}$-Fe$^{3+}$, Fe$^{3+}$-$R^{3+}$, and $R^{3+}$-$R^{3+}$).[1] The complex interplay between the 4$f$ electrons of rare earth ions and the 3$d$ electrons of iron ions gives rise to a rich spectrum of magnetic, multiferroic, and magneto-optical behaviors, making them highly attractive for applications in spintronics, information storage, and sensor technologies.[3-5] The magnetic properties of $R$FeO$_3$ are significantly influenced by the interactions between rare-earth ions and iron ions through spin-lattice coupling. Recent studies have demonstrated that strategic doping with rare-earth ions can effectively modify these interactions,[6-9] leading to tailored magnetic properties. For instance, Nd$^{3+}$ doping Dy$_{0.9}$Nd$_{0.1}$FeO$_3$ single crystal exhibits a hybrid magnetic configuration $\Gamma_{4+3}$ with magnetization along the $b$-axis and $c$-axis at elevated temperatures, which is attributed to lattice distortion and exchange interaction energy induced by Nd$^{3+}$.[6] Similarly, partial substitution of Sm with Tm in the Sm$_{0.7}$Tm$_{0.3}$FeO$_3$ single crystal enables the tuning of spin reorientation transition to lower temperature.[7] The incorporation of Yb ions into SmFeO$_3$ single crystals has been shown to effectively reduce the spin reorientation transition temperature ($T_{SR}$).[8] Furthermore, doping PrFeO$_3$ with Sm$^{3+}$ ions allows for precise control over the spin reorientation transition temperature across a wide range.[9] Despite these significant advancements, the effect of doped rare-earth ions on magnetic properties in $R$FeO$_3$ systems is still remained to be fully elucidated. A comprehensive understanding is crucial for the atomic-scale design and synthesis of $R$FeO$_3$-based materials with optimized properties for specific applications. This knowledge gap presents both a challenge and an opportunity for further research in this field.

Among the rare earth elements, the Sm$^{3+}$ magnetic moment order is highest (~150 K), which induces the spin-reorientation phase transition of SmFeO$_3$ above room temperature (~480 K), exhibiting great potential application in room-temperature devices. SmFeO$_3$ stands out among rare earth orthoferrites for its high-temperature spin reorientation, magnetization reversal, and strong spin-phonon coupling.[10-12] Due to the competition among the Fe$^{3+}$-Fe$^{3+}$, Sm$^{3+}$-Fe$^{3+}$, Sm$^{3+}$-Sm$^{3+}$, and Dzyaloshinskii-Moriya (DM) exchange interaction, the SmFeO$_3$ are weakly ferromagnetic based on the canted antiferromagnetic order below the Neel temperature.[12] In the temperature interval 680 – 480 K, Fe$^{3+}$ magnetic moment exhibits G-type antiferromagnetic order along $a$-axis, and the canted ferromagnetic moment point along $c$-axis induced by DM exchange interaction ($\Gamma_4$ phase, Fig. 1(a)). Sm$^{3+}$ magnetic moment is paramagnetic under the molecular field of Fe$^{3+}$ ions sublattice. When the temperature is close to 480 K, the Sm$^{3+}$-Fe$^{3+}$ exchange interaction is enhanced. Meantime, the direction of Fe$^{3+}$ antiferromagnetic moment rotates from $a$-axis to $c$-axis (spin-reorientation), and the canted weak ferromagnetic moment aligns along $a$-axis ($\Gamma_2$ phase, Fig. 1(b)). As the temperature decreases to 150 K, the antiferromagnetic order of Sm$^{3+}$ ion emerges along $b$-axis and the net magnetic moment of canted anti-ferromagnetism aligns along -$a$-axis. Because the sublattice magnetization of Sm$^{3+}$ and Fe$^{3+}$ ions exhibit different temperature dependence, the magnetic compensation point ($M$=0) emerges at ~3.5 K.[12]

The magnetic properties of SmFeO$_3$ are predominantly governed by the Sm$^{3+}$-Fe$^{3+}$ exchange interaction, which plays a pivotal role in determining its overall magnetic behavior. However, the practical application of SmFeO$_3$ in room-temperature spintronic devices is significantly hindered by its relatively high spin reorientation temperature. Despite extensive research on SmFeO$_3$,[8,9,13-17] the key factors influencing its magnetic properties remain incompletely understood, and a comprehensive phase diagram correlating rare-earth doping type, doping content, and magnetic properties has yet to be established. By leveraging the strong coupling between the lattice, orbital, spin, and charge degrees of freedom in strongly correlated oxides, strategic doping can effectively modulate the spin reorientation temperature and other magnetic properties of SmFeO$_3$. Such modulation not only enhance fundamental understanding of spin-lattice coupling behind tunable magnetic properties but also pave the way for its application in room-temperature spintronic devices.

In this work, the high-quality Sm$_{0.7}$Tb$_{0.3}$FeO$_3$ single crystal has been synthesized by the optical floating-zone furnace, where the novel magnetic properties have been observed. The spin-reorientation of Sm$_{0.7}$Tb$_{0.3}$FeO$_3$ single crystal appears at ~350 K. The exchange interaction between $R^{3+}$ and $Fe^{3+}$ becomes ferromagnetic coupling and the magnetic compensation point disappears at low temperature. Interestingly, the magnetic hysteresis along *a*-axis exhibits a double-triangle loop below $T_{SR}$ and then transforms to a rectangular loop above $T_{SR}$. On the contrary, the magnetic hysteresis along the *c*-axis is rectangular below $T_{SR}$. The asymmetric features of nucleation field are explained basing on modified two domains model with mean filed correction. The results of magnetic domain measurement indicate that the simplified two-domain model successfully explains the macroscopic shape of the hysteresis loop but fails to fully describe the microscopic details of domain evolution.

The measurements were performed on a 3×3×1 mm$^3$ single crystal grown by the optical floating-zone furnace technique (FZ-T-10000-H-VI-P-SH, Crystal System Corp). Detailed single crystal growth parameters are provided in the Supplementary Material. The crystal structure was measured by High-resolution X-ray diffraction (HRXRD, D8 Discover, Bruker). The magnetic properties were characterized by the physical properties measurement system (PPMS-9, Quantum Design) in the temperature range of 2 K-380 K. The external field was applied along the *a* or *c*-axis with the accuracy of 3°. A commercial AFM/MFM (Atto AFM/MFM Ixs; Attocube Systems) was used to scan the topography and magnetic images at 150 K. During the measurement, the MFM was performed in constant height mode (single pass) with PPP-MFMR tip from Nanosensors under phase mode, with the lift height is 50 nm.

The X-Ray Diffraction (XRD) 2$\theta$-$\omega$ scan of the Sm$_{0.7}$Tb$_{0.3}$FeO$_3$ single crystal along the *a*-axis direction only reveals (*l*00) diffraction peaks, as depicted in Fig. 1(c), indicating that the single crystal possesses an ideal perovskite-type structure without any impurity phases. Based on Bragg's law, the lattice constant of the *a*-axis is determined to be 5.35 Å, which lies between the *a*-axis lattice constants of SmFeO$_3$ (5.39 Å) and TbFeO$_3$ (5.30 Å) single crystals. This intermediate value can be attributed to the lanthanide contraction effect, where the ionic radius of $Tb^{3+}$ is smaller than that of $Sm^{3+}$. Consequently, the incorporation of Tb into the crystal structure results in a reduction of the lattice constant in the $Sm_{0.7}Tb_{0.3}FeO_3$ single crystal.

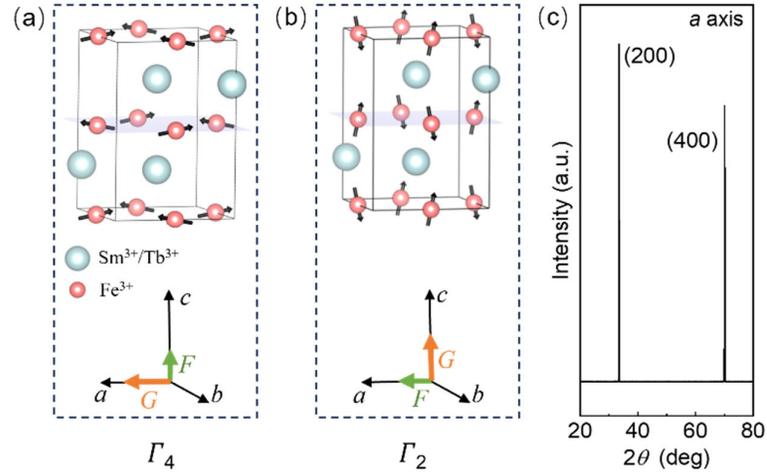

FIG. 1. The schematic magnetic structure of $SmFeO_3$ with the space group *Pbnm*, showing ideal positions of iron ions (red small balls) and rare-earth ions (blue large balls), and spin directions of iron ions for (a) $\Gamma_4$ and (b) $\Gamma_2$. (c) XRD $2\theta$-$\omega$ scan of $Sm_{0.7}Tb_{0.3}FeO_3$ single crystal along *a*-axis.

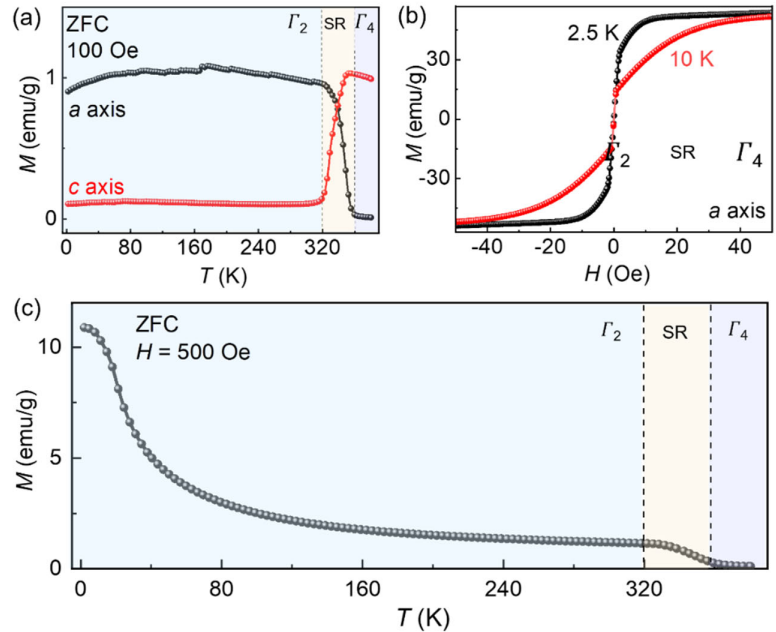

FIG. 2. (a) Zero-field-cooled $M(T)$ curves measured with 100 Oe magnetic field along *a*-axis and *c*-axis for $Sm_{0.7}Tb_{0.3}FeO_3$ single crystal. (b) *M* vs. *H* for $Sm_{0.7}Tb_{0.3}FeO_3$ single crystal measured along *a*-axis at 2.5 K and 10 K. (c) Zero-field-cooled $M(T)$ curves measured with 500 Oe magnetic field along *a*-axis for

$Sm_{0.7}Tb_{0.3}FeO_3$ single crystal.

Figure 2(a) presents the temperature-dependent magnetization (*M*) along the *a*-axis and *c*-axis for the $Sm_{0.7}Tb_{0.3}FeO_3$ single crystal. The measurement was conducted by first cooling the sample to 2 K in a zero magnetic field, followed by heating under an applied magnetic field of 100 Oe. Distinct spin reorientations are observed along both the *a*-axis and *c*-axis within the temperature range of 320-360 K. Above 360 K, the sample presents a non-collinear antiferromagnetic structure characterized by an antiferromagnetic moment aligned along the *a*-axis and a weak canted ferromagnetic moment along the *c*-axis, corresponding to the $\Gamma_4$ phase. As the temperature decreases, a spin reorientation transition ($\Gamma_4$ to $\Gamma_2$) initiates, with the magnetic moment beginning to rotate within the (*a*, *c*) plane at 360 K. By 320 K, the antiferromagnetic moment fully aligns along the *c*-axis, while the canted ferromagnetic moment reorients to the *a*-axis, marking the $\Gamma_2$ phase. The spin reorientation originates from the temperature-driven change of magnetic anisotropy of the $R^{3+}$ 4*f* electrons mediated by a small effective $R^{3+}$-$Fe^{3+}$ magnetic exchange interaction.[18,19] Besides, the different rare earth ions cause perovskite lattice distortions, altering Fe-O-Fe bond angles and lengths. This structural change modifies the superexchange interaction between $Fe^{3+}$ ions, adjusting antiferromagnetic anisotropy and spin reorientation temperature. Thus, the incorporation of Tb into $SmFeO_3$ significantly modulates the $R^{3+}$-$Fe^{3+}$ and $Fe^{3+}$-$Fe^{3+}$ exchange interaction and resulting temperature-dependent anisotropy, making a spin reorientation temperature that is lower than that of pure $SmFeO_3$ (480 K) but substantially higher than that of $TbFeO_3$ (<10 K). The spin reorientation temperature of $Sm_{0.7}Tb_{0.3}FeO_3$ single crystal is near the operating temperature of room temperature storage devices, which brings opportunities to promote its application in room temperature storage devices.

In $SmFeO_3$ single crystals, the $Sm^{3+}$-$Fe^{3+}$ exchange interaction exhibits antiferromagnetic coupling, resulting in a magnetic compensation point (*M*=0) at 3.5 K. However, in $Sm_{0.7}Tb_{0.3}FeO_3$, no compensation point is observed along the *a*-axis below 320 K (Fig. 2(a)), indicating a significant modification of the $R^{3+}$-$Fe^{3+}$ exchange interaction due to the coexistence of $Sm^{3+}$-$Fe^{3+}$ (antiferromagnetic) and $Tb^{3+}$-$Fe^{3+}$ (ferromagnetic) couplings. Notably, under a 500 Oe field, the *a*-axis magnetization increases continuously to 11 emu/g at 2 K (Fig.2(c)), with a sharp rise below 150 K coinciding with the onset of $Sm^{3+}$ magnetic ordering, indicating that the $R^{3+}$-$Fe^{3+}$ exchange interaction tends to be ferromagnetic coupling. To further investigate the $R^{3+}$-$Fe^{3+}$ exchange interaction, magnetic hysteresis measurements were conducted at low temperatures under high magnetic fields. As shown in Fig. 2(b), the *M-H* curves along *a*-axis of $Sm_{0.7}Tb_{0.3}FeO_3$ single crystal exhibit typical soft ferromagnetic hysteresis loops, distinct from the behavior previously reported for pure $SmFeO_3$ single crystals.[12] At 2.5 K, the *M-H* curve reveals the saturation magnetization of 53.78 emu/g, equivalent to 9.9 $\mu_B$ per formula unit. The magnetic moments of $Fe^{3+}$, $Sm^{3+}$ and $Tb^{3+}$ ions are 5.9, 1.5 and 9.4 $\mu_B$, respectively, and the magnetic moments of per $Sm_{0.7}Tb_{0.3}FeO_3$ unit is theoretically predicted to be 9.8 $\mu_B$.

The close agreement between the experimental and theoretical values indicates that the magnetic moments of all ions align along the external magnetic field direction at 2.5 K. At temperatures above 2.5 K, however, the 5 T magnetic field is insufficient to fully align the magnetic moments along the *a*-axis due to thermal perturbations. The ferromagnetic coupling between $Tb^{3+}$ and $Fe^{3+}$ competes with the antiferromagnetic coupling between $Sm^{3+}$ and $Fe^{3+}$, leading to a transition in the overall magnetic properties of the system from purely antiferromagnetic to partially ferromagnetic upon the incorporation of $Tb^{3+}$. This transition may result in a more complex magnetic anisotropy, which was originally dominated by antiferromagnetism. Generally, ferromagnetic materials exhibit weaker anisotropy due to their susceptibility to external magnetic fields, whereas antiferromagnetic materials typically possess stronger anisotropy. As $Tb^{3+}$ is introduced, the antiferromagnetic component weakens while the ferromagnetic component increases, potentially leading to a reduction in the overall magnetic anisotropy of the material. Consequently, the magnetic moments may more readily approach saturation under an external magnetic field. This transition underscores the significant role of $Tb^{3+}$ in modulating the magnetic interactions.

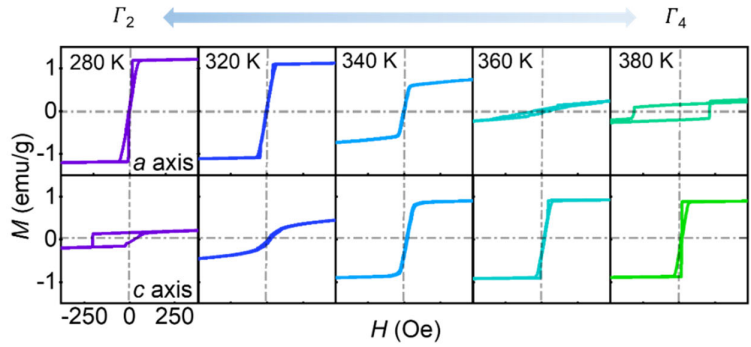

FIG. 3. *M* vs. *H* for $Sm_{0.7}Tb_{0.3}FeO_3$ single crystal measured along *a*-axis and *c*-axis with different temperatures from 280 K to 380 K.

Near the spin-reorientation temperature (Fig. 3), the *a*-axis hysteresis exhibits double-triangular loops below the transition, while the *c*-axis shows rectangular loops; above the transition, the *a*-axis loops become rectangular, and the *c*-axis loops present a double-triangular shape, reflecting the reorientation of magnetic moments from the *a*-axis to the *c*-axis. The result can be explained by considering two-domain mechanism.[20,21]

The two domains are separated by one domain wall, and the position *x* of the wall is applied to characterize the magnetic state. The total magnetic moment can be expressed as *Mx*, where *M* is the total magnetic moment of the sample along *a*-axis. The domain wall remains inside the sample for $|x|<1$, while no domain wall is in sample for $|x|>1$. The magnetic field to push the wall out of the sample is defined as the expulsion field ($H_{exp}$). The jumps of hysteresis loops correspond to the nucleation of the domain wall, named as the nucleation field ($H_{n\pm}$).

Considering both the balance of Zeeman energy and the demagnetization energy with coefficient $A$ capturing the properties of the sample shape, the magnetic energy of the sample can be described as:[20]

$$E = AM^2x^2 - MHx \tag{1}$$

In the simplest model of wall nucleation, it is assumed that there is a potential barrier $V(x)$ near $x = 1$ with a characteristic height of $V_0 \ll AM^2$ and a width of $\delta \ll 1$. When the pressure $dE/dx|_{x=1}$ exceeds the maximum value of the barrier resistance force $(dV/dx)_{max} \sim V_0/\delta$, the wall can overcome the potential barrier. The relation between the expulsion field ($H_{exp}$) and the nucleation field ($H_{n\pm}$) is given by:[20]

$$H_{exp} = 4AM \tag{2}$$

$$H_{n+} = 4AM(T) - C/M(T), \; C \sim V_0/\delta \tag{3}$$

$$H_{n-} = C/M(T) - 4AM(T) \tag{4}$$

According to the theoretical result, the magnetization process can be divided into two situation.

For $H_{n+} < -H_{exp}$, the domain wall quickly sweeps from one side, and leaves the sample at other side. Therefore, a jump from $M$ to $-M$ emerges, and the shape of magnetic hysteresis is rectangular. When the magnetization is small, it corresponds to the case of $H_{n+} < -H_{exp}$. Above the spin reorientation temperature, the hysteresis loop of the $a$-axis is rectangular, while below the spin reorientation temperature, the hysteresis loop of the $c$-axis is rectangular.

For $H_{n+} > -H_{exp}$, when the applied magnetic field decreases to $H_{n+}$, the domain wall emerges and stays in the equilibrium position, corresponding to the central branch in magnetic hysteresis loops. The shape of magnetic hysteresis is double-triangular.

As is shown in Fig. 4(a), the magnetic behavior of $Sm_{0.7}Tb_{0.3}FeO_3$ single crystal at all temperatures below $T_{SR}$ acquires a shape with "double-triangular loop", whose magnetization process is consistent with the theoretical prediction of the situation $H_{n+} > -H_{exp}$. The $H_{n\pm}$ vs. $T$ curves extracted from $M$-$H$ loops are shown in Fig. 4(b). The $|H_{n\pm}|$ decreases with the increase of temperature, corresponding to the equation (3)(4) in general.

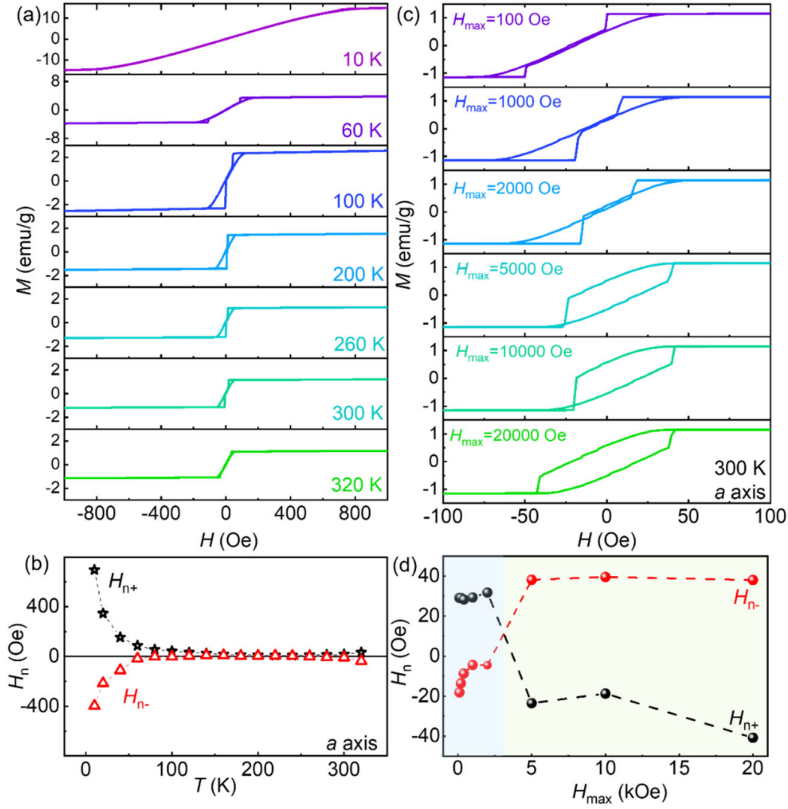

FIG. 4. (a) $M$ vs. $H$ for $Sm_{0.7}Tb_{0.3}FeO_3$ single crystal measured along $a$-axis at different temperatures in the low magnetic field region. (b) Experimental temperature dependence of the nucleation field $H_{n\pm}(T)$. (c) $M$ vs. $H$ for $Sm_{0.7}Tb_{0.3}FeO_3$ single crystal measured along $a$-axis at 300 K with different the maximum applied magnetic field ($H_{max}$). (d) Experimental $H_{max}$ dependence of the nucleation field $H_{n\pm}(T)$.

It is worth noting that the positive and negative nuclear fields $H_n$ are not same. The result is $|H_{n+}| > |H_{n-}|$, which is different from the conclusion $|H_{n+}| = |H_{n-}|$ for the equation (3)(4). The mean field method such as the molecular field correction is necessary to the previous to two domain model.

The total moment $M$ can be described as follows:

$$M = M_R + M_{Fe} \qquad (5)$$

where $M_R$ and $M_{Fe}$ are the magnetic moment of rare earth $R^{3+}$ and $Fe^{3+}$ ions, respectively. Based on Langevin functions, the magnetic moment of $Fe^{3+}$ ions can be defined as:

$$M_{Fe} = M_{Fe,S} L(x_{Fe}) \qquad (6)$$

$$L(x) = \cosh x - \frac{1}{x} \qquad (7)$$

$$x_{Fe} = \frac{(\gamma_{R-Fe}M_R + \gamma_{Fe}M_{Fe} + \mu_0 H)M_{Fe,S}}{k_B T} \quad (8)$$

Similarly, the magnetic moment of $R^{3+}$ ions can be described as:

$$M_R = \mu_R L(x_R) \quad (9)$$

$$x_R = \frac{(\gamma_{R-Fe}M_{Fe} + \mu_0 H)\mu_R}{k_B T} \quad (10)$$

In which $\gamma_{Fe}$ and $\gamma_{R-Fe}$ are molecular field constant of $Fe^{3+}$-$Fe^{3+}$ and $R^{3+}$-$Fe^{3+}$ exchange interaction, respectively. $M_{Fe,S}$ and $\mu_R$ are the saturated magnetic moment of $Fe^{3+}$ and $R^{3+}$, respectively.

When $H \gg H_C$, and the temperature is low, Langevin functions can be described as:

$$L(x) \approx \frac{x}{3} \quad (11)$$

Therefore, the total magnetic moment is as follows:

$$M = M_{Fe} + M_R = \lambda H \quad (12)$$

$$\lambda \quad (13)$$
$$= \frac{\mu_0 \mu_R^2}{3k_B T} + \left(1 + \frac{\gamma_{R-Fe}\mu_R^2}{3k_B T}\right)$$
$$* \left(\frac{3k_B T \mu_0 + \gamma_{R-Fe}\mu_R^2 \mu_0}{3k_B T(3k_B T - \gamma_{Fe}M^2_{Fe,S}) - M^2_{Fe,S}\gamma^2_{R-Fe}\mu_R^2}\right)M^2_{Fe,S}$$

According to the equation (3)(4), the nuclear field can be described as:

$$H_{n+} = \sqrt{\frac{C}{\lambda(2A\lambda - 1)}} \quad (14)$$

$$H_{n-} = -\sqrt{\frac{C}{\lambda(2A\lambda + 1)}} \quad (15)$$

Hence, when the molecular field is taken into account, we have $|H_{n+}| > |H_{n-}|$, corresponding to the experimental result.

As shown in Fig. 4(c) and 4(d), the effect of the maximum applied magnetic field on the magnetization process was investigated. As the applied magnetic field increases, the shape of the hysteresis loop transitions from a double-triangular to a rectangular form, indicating a significant change in the magnetization behavior of the material. Notably, when the maximum applied field exceeds 2000 Oe, the nucleation field ($H_n$) undergoes a sign reversal, implying that the direction of the applied field plays a decisive role in the nucleation process, leading to a change in the preferential orientation of magnetic domains during initial magnetization. This reversal is closely related to the $R^{3+}$-$Fe^{3+}$ interaction, which generates an effective field on the order of kOe[20] and critically influences the overall magnetic properties, particularly the nucleation and propagation of magnetic domains. Additionally, the nucleation field is modulated by factors such as energy barriers associated with magnetic anisotropy, and domain wall motion. These factors collectively determine the nucleation and growth of magnetic domains under different applied fields. This transition highlights the complex interplay between the applied field and switching dynamics of magnetic domain.

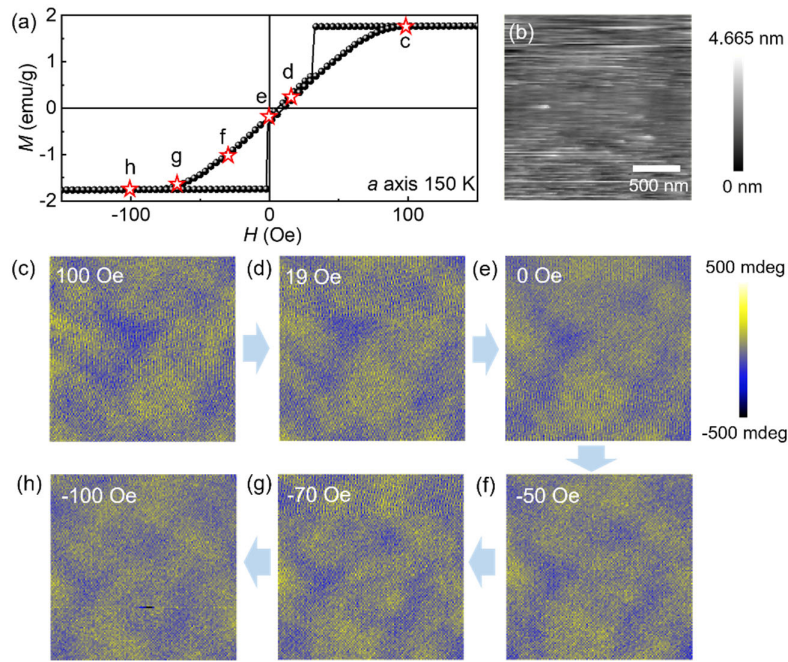

FIG. 5. (a) The magnetic hysteresis loops along $a$-axis of $Sm_{0.7}Tb_{0.3}FeO_3$ single crystal at 150 K. (b) The morphology of the $Sm_{0.7}Tb_{0.3}FeO_3$ single crystal along $a$-axis from the Atomic Force Microscope image, and (c-f) is the corresponding magnetic force microscope image at different magnetic fields.

To further investigate the magnetic behavior of $Sm_{0.7}Tb_{0.3}FeO_3$ single crystals, magnetic domain measurement was conducted. Fig. 5(a) shows the magnetic hysteresis loop at 150 K, exhibiting a typical double-triangular hysteresis loop. Fig. 5(b) presents the atomic force microscope image (AFM) topography of the $Sm_{0.7}Tb_{0.3}FeO_3$ single crystal along the $a$-axis to prevent the influence of surface morphology. Fig. 5(c-f) display the magnetic force microscopy (MFM) images along the $a$-axis under different applied magnetic fields. During the measurement process, a magnetic field of 200 Oe was initially applied to the sample, followed by a gradual reduction of the field while monitoring the evolution of the magnetic domains. In MFM image, the domain contrast is primarily determined by the canted ferromagnetic moment along the *a*-axis, indicating that the domain contrast does not directly reflect the *c*-axis antiferromagnetic order but rather the behavior of the *a*-axis parasitic ferromagnetism.

According to the results in Fig. 5(c-f), as the magnetic field is decreased from 100 Oe to 0 Oe, the canted ferromagnetic component along the +*a* axis gradually diminishes, leading to a reduction in the blue domain region and a decrease in contrast. Conversely, when the magnetic field increases from 0 Oe to -100 Oe, the canted ferromagnetic component along the -*a* axis gradually increases, resulting in the expansion of the yellow domain region. The above process demonstrates that the domain evolution is mainly driven by changes in the canted ferromagnetic component along the *a*-axis of the antiferromagnetic moment, accompanied by significant domain wall motion. By correlating these results with the hysteresis loop, we speculate that the jumps in the double-triangular hysteresis loop primarily correspond to domain wall motion, while the linear regions following the jumps are more likely related to the nonlinear response of the canted ferromagnetic component along the *a*-axis. Although the simplified stripe double-domain model mentioned earlier successfully explains the macroscopic shape of the hysteresis loop and the asymmetric nucleation behavior, it does not fully describe the microscopic details of domain evolution. This suggests that further research based on magnetic structure characterization and theoretical model improvements is necessary to reveal the mechanism behind domain switching dynamics and resultant magnetization behavior in non-collinear antiferromagnetic materials.

In conclusion, this study demonstrates that doping of rare-earth perovskite SmFeO$_3$ compounds with Tb$^{3+}$ ions can significantly modify their magnetic properties and modulate the spin reorientation temperature to approximately 350 K, making them promising candidates for room-temperature spintronic applications. The absence of a magnetic compensation point at low temperatures and the full saturation of both $R^{3+}$ and Fe$^{3+}$ ion moments under high magnetic fields indicate that Tb$^{3+}$ doping enhances ferromagnetic coupling between the $R^{3+}$ and Fe$^{3+}$ sublattices. Furthermore, the transformation of the hysteresis loop along the *a*-axis from a double triangular shape below the spin reorientation temperature to a rectangular shape above it. The nucleation field's dependence on temperature and applied magnetic field, explained by a simplified two-domain model, underscores the role of domal wall movement in orthoferrites. The findings from magnetic domain measurement suggest that the proposed two-domain model provides a good explanation for the macroscopic characteristics of the hysteresis loop, but it is inadequate in fully representing the detailed microscopic processes of domain evolution. This study lays the foundation for establishing a comprehensive phase diagram of rare-earth doping types, doping concentrations, and magnetic properties, while successfully optimizing the performance of doped SmFeO$_3$ for next-generation spintronic technologies.

This work was supported by the National Natural Science Foundation of China

## AUTHOR DECLARATIONS
**Conflict of Interest**
The authors have no conflicts to disclose.

**Author Contributions**
**Mingzhu Xue:** Data curation (equal); Conceptualization (equal); Formal analysis (equal); Investigation (lead); Writing-original draft (lead); Writing-review & editing (equal). **Xin Li:** Conceptualization (equal); Data curation (equal); Formal analysis (equal); Writing-review & editing (equal). **Shilei Ding:** Conceptualization (equal); Investigation (equal); Writing-review & editing (equal). **Qixin Li:** Resources (equal); Investigation (equal); Writing-review & editing (equal). **Wenhao Di:** Resources (equal); Investigation (equal); Writing-review & editing (equal). **Anhua Wu:** Resources (equal); Investigation (equal); Writing-review & editing (equal), Funding acquisition (equal). **Bin He:** Investigation (equal); Writing-review & editing (equal). **Shishen Yan:** Investigation (equal); Writing-review & editing (equal). **Wenyun Yang:** Conceptualization (equal), Formal analysis (equal), Funding acquisition (equal), Project administration (lead), Supervision (lead), Writing – review & editing (equal). **Jinbo Yang:** Conceptualization (equal); Investigation (equal); Writing-review & editing (equal).

## DATA AVAILABILITY
The data that support the findings of this study are available from the corresponding author upon reasonable request.